%% file: main_prelim_arx.tex
\documentclass[12pt,fleqn]{article}
\usepackage[utf8]{inputenc}
\usepackage{a4wide}

\usepackage{amsmath}
\usepackage[round]{natbib}
\usepackage{graphicx}   
\usepackage{mathtools}
\usepackage{bm}
\usepackage{appendix}
\usepackage{amsthm}

\usepackage{xcolor}
\usepackage{ulem}
\usepackage{amsmath}
\usepackage{amssymb}
\usepackage{rotating}
\usepackage{comment}
\usepackage[hidelinks]{hyperref}
\hypersetup{
    colorlinks,
    linkcolor={red!50!black},
    citecolor={blue!50!black},
    urlcolor={blue!80!black}
}

\usepackage{array}

\begin{document}

\title{Faster estimation of the Knorr-Held Type IV space-time model}
\author{Fredrik Lohne Aanes\thanks{Email: fredriklohneaanes@gmail.com}\\
Norwegian Computing Center\and 
Geir Storvik\thanks{Email: geirs@math.uio.no, corresponding author}\\
University of Oslo and Norwegian Computing Center }
\maketitle

\bigskip

\input{main_body.tex}

\end{document}

%% file: main_body.tex
\begin{abstract}
In this paper we study the type IV Knorr Held space time models. Such models typically apply intrinsic Markov random fields and constraints are imposed for identifiability. INLA is an efficient inference tool for such models where constraints are dealt with through a conditioning by kriging approach.
When the number of spatial and/or temporal time points become large, it becomes computationally expensive to fit such models, partly due to the number of constraints involved. 
We propose a new approach, HyMiK, dividing constraints into two separate sets where one part is treated through a mixed effect approach while the other one is approached by the standard conditioning by kriging method, resulting in a more efficient procedure for dealing with constraints. The new approach is easy to apply based on existing implementations of INLA.

We run the model on simulated data, on a real data set containing dengue fever cases in Brazil and another real data set of confirmed positive test cases of Covid-19 in the counties of Norway. For all cases we get very similar results when comparing the new approach with the tradition one while at the same time obtaining a significant increase in computational speed, varying on a factor from 2 to 4, depending on the sizes of the data sets.

\end{abstract}

\noindent%
\vfill

 
 \section{Introduction}

The availability of temporal areal data has greatly improved in recent years and Bayesian analysis have gained increasing interest for such data. 
Temporal and spatial dependence which are not accounted for by available explanatory variables are typically dealt with through random effects with specific structures. 
\citet{knorr2000bayesian} introduced the Knorr-Held separable space-time models for intrinsic Gaussian Markov random fields. These  (and similar) models have become highly popular within Bayesian disease mapping~\citep{lawson2018bayesian}, analysis of mortality rates~\citep{khana2018bayesian},
analysis of environmental systems~\citep{wikle2003hierarchical},
election forecasts~\citep{pavia2008election} and
social sciences~\citep{shoesmith2013space,vicente2020bayesian,williams2019spatiotemporal} as well as in many other applications. 

For the (Bayesian) analysis of spatio-temporal areal data, there is a huge computational challenge. This is particular the case for the Knorr-Held models. Methods and generic packages such as CARBayesST~\citep{lee2018spatio} for Markov chain Monte Carlo (MCMC) are available, but can be inherently slow for large data sets. 
Integrated nested Laplace approximation \citep[INLA,][]{rue2009approximate} is very often used to fit such Bayesian models. In~\cite{nazia2022methods}, a review over different techniques used to model Covid-19, 14 out of the 18 Bayesian studies used the INLA approach. INLA uses a combination of analytical approximations and numerical algorithms to approximate the marginal posterior distributions. One can then avoid using the traditional MCMC sampling method to obtain inference. However, handling a large number constraints, typically needed for the Knorr-Held models, is a challenge also within the INLA framework.

There are four different Knorr-Held models, where the complexity of the model depends on the structure for the interaction term. We focus on the type IV-model. It is the most complex model as it assumes the precision matrix of an interaction between space and time term is both spatially and temporally structured. An additional complexity is due to that the random effects typically are modeled through intrinsic Gaussian Markov random fields which are improper Gaussian models (singular precision matrices). Typically this also leads to non-identifiable models. 
Additional identifiability constraints are needed to make the model identifiable. 

To make the model estimable, \citet{schrodle2011spatio} suggest to make the Knorr-Held term proper by imposing several constraints on the prior.
As an alternative, \citet{goicoa2018spatio},  suggest to make the posterior proper, only including enough constraints so that the model is identifiable. Several constraints are needed when using either sets of constraints, although the latter will have somewhat fewer constraints. The high number of constraints needed make the models hard to compute for large studies (where either $n_T$, the number of time steps, and/or $n_S$, the number of spatial regions, is very large). 
The latest version of INLA has been made much more efficient by better use of parallelism and state-of-the-art sparse linear solvers~\citep{gaedke2023parallelized,VANNIEKERK2023107692}, but still require substantial computational resources.

Incorporation of constraints can be performed in different ways. Conditioning by kriging \citep{cressie1993statistics} is a post-processing procedure which "corrects" unconstrained random processes into constrained ones through a linear transformation. This is done in each iteration of the estimation procedure.
The computational cost of constraints when using this approach is $\mathcal{O}(nk^2)$ where $k$ the number of constraints and $n$ is the dimension of the field~\citep[see e.g][]{bakka2018spatial}.
Conditioning by kriging is the default procedure within INLA. 

An alternative approach is to formulate the model as a mixed effect model with the inclusion of an appropriate design matrix that accounts for the constraints. No post-processing is then needed, but the resulting precision matrices are more dense. This approach can be implemented through the use of the
"z-model" in INLA.
In this paper we show that by \emph{combining} conditioning by kriging with the mixed effect model in a suitable way, significant computational efficiency can be gained for the Knorr-Held model type IV term. A subset of the constraints will be dealt with through the mixed effect model, 
while the remaining ones use the conditioning by kriging step approach.  By suitable constructions of such subsets, the increasing computational effort due to the mixed effect model will be much less than the decreasing computational effort due to the number of constraints that needs to be dealt with through the conditioning by kriging step.
This holds irrespective of the computational resources. 

Recently {\cite{fattah2022approximate}} proposed a new way of estimating constrained intrinsic Gaussian Markov random fields. They utilize the correspondence between the constraints suggested by \cite{schrodle2011spatio} and the Moore-Penrose inverse of the precision matrix. The procedure is still computationally demanding, but 40 fold faster inference can be obtained paralleling the computational efforts on a large server. They run it on 25 nodes Cascade
Lake nodes, each with 40 cores, 2.50 GHz, 384 GB/usable 350 GB. We do not have the same computational resources available, and cannot run the models equally fast. We therefore do not try to compare our approach with the {\cite{fattah2022approximate}} approach. However, as we will see, we can obtain similar speed-ups with our method without the need for numerous cores.

The paper is structured as follows. In Section 
\ref{sec:model_formulation} we formally introduce space-time models
and discuss how constrained fields are estimated in INLA. In Section \ref{sec:projection method for inference} we present the new method of estimating constrained fields. In Section \ref{sec:numstud} we study three datasets, one simulated and two real data sets. We also discuss the results. Finally,  Section~\ref{section:Discussion}, provide a summary and a discussion.

\section{Model formulation}{\label{sec:model_formulation}

We consider a modeling framework where we have data $\{y_{t,s},t=1,...,n_T,s=1,...,n_S\}$ corresponding to observations at time $t$ within area/region $s$. The observations are assumed conditionally independent with  
\begin{subequations}\label{eq:model}
\begin{align}
y_{t,s}|\eta_{t,s}\sim& p_y(y_{t,s}|\eta_{t,s},\phi)\\
\eta_{t,s}=&\mu+\bm u_{t,s}^T\bm\beta+\alpha_t+\gamma_s+\delta_{t,s}=(\mathcal{A}\bm x)_{t,s}\label{eq:model.eta}
\end{align}
\end{subequations}
for $t=1,...,{n_T},s=1,...,{n_S}$. Here, $p_y(\cdot)$ is a distribution within the exponential family with
$E[y_{t,s}]=g^{-1}(\eta_{t,s})$ for some appropriate link function $g(\cdot)$ while $\phi$ is a dispersion parameter. The $\mathcal{A}$-matrix is a sparse design matrix, while $\bm x=\{\mu, \bm\beta,\bm \alpha,\bm \gamma,\bm\delta\}$ is the latent field.   
The intercept $\mu$ and the regression coefficients $\bm\beta$ are treated as fixed effects, the latter related to covariates $\bm u_{t,s}$\footnote{We have used the somewhat unusual notation $\bm u$ for the covariates here in order to follow the standard notation of $\bm x$ for latent variables later on.}.
The two main random effects $\{\alpha_t\}$ (temporal component)
and  $\{\gamma_s\}$ (spatial component) are both assumed to be 
zero mean Gaussian Markov random fields with  precision matrices $\bm Q_{\alpha}$ and $\bm Q_{\gamma}$, respectively.  Also the interaction term  $\{\delta_{t,s}\}$ is assumed to following a zero mean Gaussian distribution, with precision matrix $\bm Q_{\delta}$. Typically, the precision matrices for the random effects are singular, corresponding to \emph{intrinsic} Gaussian Markov random field models. In such settings, the model can become non-identifiable and constraints are in such cases required on the random effects.

 For the temporal component, typical model choices are the random walk of order 1 (RW(1)), in which case $\alpha_t-\alpha_{n_T-1}\sim N(0,\tau_\alpha^{-1})$ or the
random walk of order 2 (RW(2)), in which case $(\alpha_t-\alpha_{n_T-1})-(\alpha_{n_T-1}-\alpha_{t-2})\sim N(0,\tau_\alpha^{-1})$. For the RW(1) model, a sum to zero constraint $\sum_{t=1}^{{n_T}}\alpha_t=0$ will make the model proper. For the RW(2) model, the additional constraint
$\sum_{t=1}^{{n_T}}t\alpha_t=0$ results in a proper model. Note however that depending on which fixed effects included, the $\{\alpha_t\}$ process might be \emph{identifiable} without the latter constraint. If a linear temporal effect is included as a fixed effect, the additional constraint needs to be included.

The \emph{intrinsic conditional autoregressive model} (ICAR) was introduced by \cite{besag1974spatial}  as a two-dimensional generalization of the  discrete penalty smoother
by \cite{whittaker1922new} and has been a popular choice for the spatial component.
Here we assume
\begin{align}\label{eq:ICAR}
\gamma_s|\bm\gamma_{-s}\sim N\left(\gamma_s;{|\mathcal{N}_s|}^{-1}\sum_{s'\in\mathcal{N}_s}\gamma_{s'},(\tau_{\gamma}|\mathcal{N}_s|)^{-1}\right),
\end{align}
where $\mathcal{N}_s$ is the set of nodes that are  neighbors to node $s$ and $|\mathcal{N}_s|$ is the number of neighbors for node $s$. 
In this case, $\bm Q_{\gamma}$ will have $\tau_{\gamma}|\mathcal{N}_s|$ on the diagonals, $-\tau_{\gamma}$ for all off-diagonals corresponding to neighbor nodes and zero otherwise.
Assuming all nodes are connected through a path of neighbor nodes, $\bm Q_{\gamma}$ has rank deficiency 1 and a typical constraint imposed is $\sum_{s=1}^{n_S}\gamma_s=0$. 

Note that for the models discussed above, we have $\bm Q_\alpha=\tau_\alpha\bm R_\alpha$ and $\bm Q_\gamma=\tau_\gamma\bm R_\gamma$ where $\bm R_\alpha$ and $\bm R_\gamma$ are known (singular) matrices, typically denoted as the structural matrices. In more general settings, some unknown parameters might be involved in the structural matrices as well, although we will not consider such cases in this paper.

\subsection{The space time interaction and corresponding constraints}
 \label{sec:spaceTime}
  We will in this section discuss the constraints needed for the spatio-temporal interaction terms $\{\delta_{t,s}\}$.
\citet{knorr2000bayesian} considered four types of interaction models corresponding to unstructured/structured precision matrices for the temporal/spatial components of $\bm Q_{\delta}$ where unstructured means independence while structured corresponds to the RW(1/2) and to the ICAR models discussed in the previous section. We will here focus on Type IV interactions where both the temporal and the spatial parts are structured with $\bm Q_\delta=\tau_\delta\bm R_{\delta}$ and
\begin{equation}\label{eq:Q.delta}
\bm R_{\delta}=\bm R_{\alpha}\otimes\bm R_{\gamma}.
\end{equation}
Denote the rank deficiencies for $\bm R_{\alpha}$ and $\bm R_{\gamma}$ by $k_{\alpha}$ and $k_{\gamma}$, respectively. Then the rank deficiency for $\bm R_{\delta}$ is  $k_{\delta}=k_{\alpha}*{n_S}+k_{\gamma}*{n_T}-k_{\alpha}*k_{\gamma}$. In order to make the distribution for $\bm\delta=(\delta_{1,1},...,\delta_{1,{n_S}},\delta_{2,1},...,\delta_{{n_T},{n_S}})$ a proper distribution, $k_{\delta}$ constraints are needed. However, in order to make the model \emph{identifiable}, fewer constraints might be needed. For instance, if the model is
\[
\eta_{t,s}=\delta_{t,s}
\]
so no fixed effects and no other random effects, all the $\delta_{t,s}$'s will be identifiable (given that there is one observation for each $(t,s)$ combination).  In practice,  at least an intercept, a temporal $\alpha_t$ as well as a spatial $\gamma_s$ effect are typically included for ease of interpretation. In these situations, constraints are needed.

 Several sets of constraints have been suggested in the literature. In the following, we will always assume that the intercept $\mu$ is included as a fixed effect, in which case the sum constraints
 \[
 \bm 1_{n_S}^T\bm\gamma=0,\quad \bm 1_{n_T}^T\bm\alpha=0
 \]
 are included in order to avoid identifiability issues between the intercept and the main effects. Here, $\bm 1_{n}$ is a vector of length $n$ with all components equal to 1.

 For the interaction term, more care is needed with respect to design of constraints, due to that it can both have identifiability issues related to the fixed effects and to the main random effects.
 The constraints suggested by \cite{goicoa2018spatio} for the interaction term, named the GC-constraints in the following, are given in the first row of Table~\ref{table:Constraints}. These constraints are based on that main random effects always are included and focus on constraining effects that are non-identifiable regardless of data available. Due to missing data (e.g no observations for a specific region) some other effects might also be non-indentifiable. We will however only consider cases where the remaining effects \emph{are} identifiable, typically based on that there is at least one observation for each combination $(t,s)$.  
 We note that the GC-constraint system is overdetermined as the rank of the constraint matrix is less than the number of constraints. 
 In practice, a set of constraints of full rank can be obtained by removing one row in 
  either
 $\bm I_{n_T}\otimes\bm 1_{n_S}^T$ or
 $\bm 1_{n_T}^T\otimes\bm I_{n_S}$.
%

\cite{schrodle2011spatio} suggest to use constraints that give a prior that is \emph{proper}, a stronger requirement than the identifiability constraint. For some settings a proper prior might be desirable, although, as noted by~\citet{goicoa2018spatio}, the extra constraints in the SC sets are not needed from an identifiability point of view unless a common linear trend and area specific linear trends are included in the model as fixed effects.  
 The SC constraints are given in the second row of Table \ref{table:Constraints}. Also this set of constraints are overdetermined and deleting one row from either
 $\bm I_{n_T}\otimes\bm 1_{n_S}^T$ or from
 $\bm 1_{n_T}^T\otimes\bm I_{n_S}$ in addition to removing one row from $\bm d^T\otimes\bm I_{n_S}$ will give a full rank set.
 Assuming an RW(2) model for the temporal structure and proper priors also are wanted for the main effects, the constraint
$\bm d^T\bm\alpha=0$
should additionally be imposed.


Depending on what fixed effects are included, additional constraints might be needed. For instance, if a linear effect in time is part of the fixed effects, the constraint $\bm d^T\bm\alpha=0$ should always be included.  Note that for both constraint sets, a sum to zero condition is implied, although not explicitly listed.

The SC constraints in Table~\ref{table:Constraints} can be seen as defined within the space spanned by the eigenvectors corresponding to the eigenvalues equal to zero for $\bm Q_\delta$. 
\citet{schrodle2011spatio} more generally suggest defining the constraint matrix $\bm A$ by those eigenvectors of $\bm R_\delta$ which span the null space. 
\citet{fattah2022approximate} considered 
this approach, using the Moore-Penrose pseudo-inverse of the prior precision matrix $\bm Q$ to obtain such constraints. 
Typically, however, the resulting \emph{covariance matrix} will be dense resulting in high computational complexity. \citet{fattah2022approximate}  took advantage of multi-core architectures and abundance of memory to deal with the computational burden, though still resulting in a computational demanding procedure.

\begin{table}
\begin{center}
\begin{tabular}{|l l|} 
 \hline
Constraint name & Constraint\\
 \hline
 Goicoa  (GC) & $[\bm I_{n_T} \otimes \bm 1_{n_S}^T] \bm\delta=\bm 0_{n_T},[\bm 1_{n_T}^T \otimes \bm I_{n_S}] \bm\delta=\bm 0_{n_S},$\\
 Schrödle (SC)& $[\bm I_{n_T} \otimes \bm 1_{n_S}^T] \bm\delta=\bm 0_{n_T},[\bm 1_{n_T}^T \otimes \bm I_{n_S}] \bm\delta=\bm 0_{n_S},$
 $[\bm d^T\otimes \bm I_{n_S}]\bm\delta=\bm 0_{n_S}$\\  \hline
\end{tabular}
\end{center}
\caption{Constraints on the interaction terms in the space-time model with smooth main spatial effect $\bm\gamma$ smooth main temporal effect $\bm\alpha$ and smooth interaction effect $\bm\delta$. 
Here $\bm d=(1,2,...,n_T)^T$. 
}  
\label{table:Constraints}
\end{table}

In general, precision matrices involved will be sparse.
This can be exploited by estimation algorithms to obtain faster inference. However, including constraints will affect the sparseness of the precision matrix for $\bm x$ and
algorithms have to correct for the constraints ways such that fast inference is still achievable. 
We will consider a Bayesian approach where we assume there exist priors for the parameters involved. 
The form of these priors will be specified within the numerical studies in Section~\ref{sec:numstud}, but a main restriction is that the parameters related to the fixed effects follow a (possible noninformative) Gaussian distribution. The constraining approach to be presented in Section~\ref{sec:projection method for inference} should in principle work for any type of prior for the remaining parameters (mainly scaling factors for the precision matrices involved).

\subsection{Integrated nested Laplace approximation}
\label{sec:LaplaceApproximation}
Integrated nested Laplace approximation  \cite[INLA,][]{rue2009approximate} has been a popular method for performing inference
within latent Gaussian models such as model~\eqref{eq:model}. 
We provide here a rough overview of the method. 
Let $\bm\theta$ be the hyperparameters involved (typically precision and dispersion parameters)  and $\bm x$ be the set of latent variables, including both random effects, the intercept and regression coefficients.
INLA starts by making an approximation
\begin{equation}\label{eq:approx.inla}
\tilde p(\bm\theta|\bm y)\propto \frac{p(\bm x,\bm\theta,\bm y)}{\tilde p_G(\bm x|\bm\theta,\bm y)}\bigg|_{\bm x=\bm x^*(\bm\theta)}
\end{equation}
where $\tilde p_G(\bm x|\bm\theta,\bm y)$ is a Gaussian approximation to $p(\bm x|\bm\theta,\bm y)$ and $\bm x^*(\bm\theta)$ is the mode of $p(\bm x|\bm\theta,\bm y)$ for a given $\bm\theta$. Note that the Gaussian prior on $\bm x$ make such an approximation reasonable.  The marginal distributions $p(x_i|\bm y)$ are then approximated by
\[
\tilde p(x_i|\bm y)=\sum_k\tilde p(x_i|\bm\theta_k,\bm y)\tilde p(\bm\theta_k|\bm y)\Delta_k
\]
where $\tilde p(x_i|\bm\theta_k,\bm y)$ is an approximation of $p(x_i|\bm\theta_k,\bm y)$ \citep[considered three alternatives with a Gaussian approximation being the simplest one]{rue2009approximate}. The $\bm\theta_k$ with their associated weights $\Delta_k$ are carefully distributed around the mode of $\tilde p(\bm\theta|\bm y)$, see~\citet{rue2009approximate} for details.

The main computational cost is related to the Gaussian approximation  $\tilde p_G(\bm x|\bm\theta,\bm y)$. Due to the Gaussian assumption on $\bm x$ and the conditional independence assumption on the observations, we have 
\begin{align}\label{eq:Gauss.approx}
p(\bm x|\bm y,\bm\theta)\propto \exp\left(-\frac{1}{2}\bm x^T\bm Q\bm x+\sum_{t,s}g_{t,s}(\mathcal{A}\bm x)\right),  
\end{align}
where $\mathcal{A}\bm x$ is specified through \eqref{eq:model.eta}. Further, $\bm Q$ is the full precision matrix for all the elements in $\bm x$. The sparsity structures of the precision matrices for the different random and fixed effects in $\bm x$ will be inherited into $\bm Q$.
The Gaussian approximations involved in INLA is based on Taylor expansions of the $g_{t,s}(\bm(\mathcal A\bm x)_{t,s})$ terms. 
 The Gaussian approximation is in particular obtained by the Taylor approximation
\begin{equation}
g_{t,s}((\mathcal{A}\bm x^{m})_{t,s})\approx g_{t,s}(\mathcal A \bm x^{m})_{t,s}+b_{t,s}(\mathcal{A}\bm x^{m})_{t,s}-\tfrac{1}{2}c_{t,s}(\mathcal A \bm x^{m})_{t,s}^2\label{eq:g.approx}
\end{equation}
based on the mode $\bm x^{m}$ of $p(\bm x|\bm\theta,\bm y)$. The mode is computed iteratively using a Newton-Raphson algorithm. The computational steps utilize the Cholesky decomposition of $\bm Q$ which is sparse if $\bm Q$ is sparse. 
 See~\citet{VANNIEKERK2023107692} for further details on the implementation and computational cost related to INLA.

\subsection{INLA and constraints}\label{sec:inf.constr}
As discussed in Section~\ref{sec:spaceTime}, (linear) constraints are usually needed in order to make intrinsic models identifiable.
There are different ways in which such constraints can be incorporated into the inferential procedure.
In case of constraints  $\bm A\bm x=\bm 0$, the density to be considered is $p(\bm x|\bm A\bm x=\bm 0,\bm y,\bm\theta)$.

The most standard way of incorporating constraints within INLA is  based on the conditioning by kriging technique~\citep[sec 2.3.3]{rue2005Gaussian} using the posterior distribution. First, denote by $\widetilde{\bm Q}$ the precision matrix corresponding to the Gaussian approximation~\eqref{eq:Gauss.approx}. Assume  $\bm x^*\sim N(\bm\nu,\widetilde{\bm Q}_\epsilon^{-1})$, where 
\begin{equation}
\label{eq:Q_eps}
  \widetilde{\bm  Q}_\varepsilon= \widetilde{\bm Q}+\bm I\varepsilon
\end{equation}
where $\varepsilon>0$ is a small number included for numerical reasons. 
Then
\begin{equation}\label{eq:cond.krig}
\bm x=\bm x^*-\widetilde{\bm Q}_\epsilon^{-1}\bm A^T(\bm A\widetilde{\bm Q}_\epsilon^{-1}\bm A^T)^{-1}\bm A\bm x^*
=[\bm I-\widetilde{\bm Q}_\epsilon^{-1}\bm A^T(\bm A\widetilde{\bm Q}_\epsilon^{-1}\bm A^T)^{-1}\bm A]\bm x^*
\end{equation}
will have the right conditional distribution~\citep[][sec 2.3.3]{rue2005Gaussian}.
This relationship is utilized in the optimization step by correcting the solution at each iteration
based on the relation
using~\eqref{eq:cond.krig} based on
\begin{equation}
\label{eq:LikelihoodEvalKriging}
    p(\bm x| \bm A\bm x=\bm 0,\bm y)=\frac{p(\bm A\bm x| \bm x,\bm y)p(\bm x| \bm y)}{p(\bm A\bm x| \bm y)}I(\bm A\bm x=\bm 0)
\end{equation}
which is evaluated at the $\bm x$ derived by~\eqref{eq:cond.krig}. Due to the additional $\varepsilon$ term, the conditioning by kriging correction will not give exact results, although the effect should be small.

 This correction step is done in each iteration of the INLA procedure.
The additional cost of this correction is $\mathcal{O}(nk^2)$ where $k$ is the number of constraints. For small $k$, conditioning by kriging is very efficient. However, for spatio-temporal random effects $k$ is quite large.

When intrinsic models are used, it is common to add a small value $\varepsilon$ to the precision matrices involved in order to make the models proper.  
If the distribution is proper, adding the value $\varepsilon$ to the diagonal should only give marginal modifications of the distribution. 
The eigendecomposition of $\bm Q_\varepsilon^{-1}=(\bm Q+\varepsilon\bm I)^{-1}$ is given by
\begin{equation}
\label{eq:Q_inv}
    \bm Q^{-1}_\varepsilon=\sum_{i=1}^n \frac{1}{\varepsilon+\lambda_i}\bm a\bm a_i^T
\end{equation}
where $\{\bm a_i\}$ are the eigenvectors of $\bm Q$ with corresponding eigenvalues $\{\lambda_i\}$.
Note that for eigenvalues $\lambda_i=0$, the additional $\varepsilon$ terms remove the problems of calculating inverses in~\eqref{eq:cond.krig} but can be numerically unstable due to the resulting $\tfrac{1}{\varepsilon}$ terms in~\eqref{eq:Q_inv}. We have a similar situation for the posterior precision $\tilde{\bm Q}$.



\section{The hybrid mixed effect and kriging method for constraining intrinsic Gaussian models}
\label{sec:projection method for inference}

In this section we propose a new way of dealing with constraints for the spatio-temporal interaction terms $\bm\delta$.
We will first demonstrate how constraints can be dealt with through the mixed effect model approach. This procedure will in itself not be very efficient, but we will then propose a new method where we combine the mixed effect approach with the conditioning by kriging approach in order to obtain a computationally efficient procedure.

\subsection{The mixed effect constraining method for Gaussian models}
The trick in the mixed effect model approach is to reformulate the model by first assuming $\bm x^*$ has precision matrix $\bm Q_\varepsilon$ and then define
\begin{equation}\label{eq:mix}
\bm \eta = \bm U\bm \beta + \bm Z \bm x^*,
 \end{equation}
 for an appropriate design matrix $\bm Z$ such that the constraints are directly incorporated into $\bm Z\bm x^*$. Assume that the constraints are given by
 \[
\bm a_i^T\bm x=0, i=n-k+1,...,n
\]
where now the $\bm a_i$'s are orthonormal eigenvectors of the prior precision matrix $\bm Q$, corresponding to eigenvalues equal to zero. Then the condition by kriging transformation~\eqref{eq:cond.krig}
reduces to 
\[
\bm x=\bm x^*-\sum_{i=n-k+1}^n\bm a_i\bm a_i^T\bm x^*
=[\bm I-\sum_{i=n-k+1}^n\bm a_i\bm a_i^T]\bm x^*
\]
so by choosing $\bm Z=\bm I-\sum_{i=n-k+1}^n\bm a_i\bm a_i^T$ we directly obtain the desired constraints. 
Note that $\bm Z^2=\bm Z$ showing that $\bm Z$ is a projection matrix. 

In INLA, the use of mixed effect models is implemented through the  "z-model"~\citep{INLA-z}, but has, to our knowledge, not been used for incorporating constraints.
The method is implemented by working with the full precision matrix for both $\bm x^*$ and $\bm x=\bm Z\bm x^*$. Even though $\bm x$ is a deterministic function of $\bm x^*$,  
for numerical reasons, it is assumed that  $\bm x|\bm x^*\sim N(\bm Z\bm x^*,\kappa^{-1}\bm I)$ for a large precision value $\kappa$ in addition to that $\bm Q$ is replaced by $\bm Q_{\varepsilon}$. This results in the joint distribution for $(\bm  x,\bm x^*)$ being a zero-mean GMRF with joint precision matrix
\begin{equation}\label{eq:Q.J}
\bm Q_J =
\begin{bmatrix}
\kappa\bm I & -\kappa\bm Z\\
-\kappa\bm Z & \kappa\bm Z + \bm Q_\varepsilon
\end{bmatrix}.
\end{equation}
The model~\eqref{eq:mix} is then related to this extended random vector through $\bm\eta=\bm U\bm\beta+\bm x$. 
In contrast to the kriging approach discussed in Section~\ref{sec:inf.constr}, the constraints are now corrected for in the prior distribution so that it follows a constrained distribution \emph{before} performing the Laplace approximation.

\subsection{The hybrid mixed effect and kriging method}
In general, the $\bm Z$ matrix can be quite dense, making the use of the mixed effect model approach computationally inefficient.
Our suggested approach is based on \emph{combining} the mixed effect approach with the conditioning by kriging approach by dividing the set of constraints into two parts. A first set of constraints, $\bm A_1\bm x=\bm 0$ with the $k_1$ rows being orthonormal are dealt with through the mixed effect model approach. The remaining constraints $\bm A_2\bm x=\bm 0$ are dealt with through the conditioning by kriging procedure. For $\bm A_2$, the rows do not need to be orthonormal. 
Note that through the mixed effect formulation~\eqref{eq:mix}, the remaining constraints are put on $\bm x=\bm Z\bm x^*$ and not on $\bm x^*$. 
This procedure, which we name \emph{HYbrid MIxed effect and Kriging method for constraining intrinsic Gaussian models} (HyMiK), results in fewer constraints needed to be corrected for through the conditioning by kriging equation~\eqref{eq:cond.krig} at the cost of a larger precision matrix $\bm Q_J$. However, if $\bm Q_J$ is sparse, the increased cost of the larger latent field can be minor compared to the reduction in computing time of the conditioning by kriging step. 

In addition to computational gain, we also anticipate that the new method will give more correct parameter estimates when using non-Gaussian observation models. The reason is that $\bm x$ has been corrected for several constraints before performing the approximation~\eqref{eq:approx.inla}. Due to the intrinsic structures of the precision matrices involved, some directions in the $\bm x$ space will have large variances, giving sources for numerical instability. When some constraints are accounted for prior to this approximation, less directions will also have this problem with larger variances.
Also, less constraints are needed to be corrected for through the conditioning by kriging approach. 
Due to that the Gaussian approximation within the HyMiK approach is performed on a much more restricted space, we expect this approximation to be more accurate.
Since the approximation is more correct, the kriging  step should also be more accurate. For some models, the kriging step of HyMiK is not needed so that no post-processing is performed. Therefore HyMiK is better than using the traditional conditioning by kriging method.  

The split of the constraints should be performed such that both the joint precision matrix $\bm Q_J$ in~\eqref{eq:Q.J} be reasonable sparse and the remaining constraints used in the conditioning by kriging procedure is low in number.
From the GC constraints listed in Table~\ref{table:Constraints}, there are two obvious choices. Either
$\bm A_1=\bm A_{S,1}\equiv\bm e_{n_T}^T\otimes\bm I_{n_S}$ or
$\bm A_1=\bm A_{T,1}\equiv\bm I_{n_T}\otimes\bm e_{n_S}^T$ can be used. Here $\bm e_n=\bm 1_n/\sqrt{n}$ is the normalized version of $\bm 1_n$. Which constraint matrix to choose will depend on the size of $n_T$ compared to $n_S$. If $n_S>n_T$, the first choice is the best one while the second one is better when $n_T>n_S$. 

Considering the choice $\bm A_1=\bm A_{S,1}$, there are $n_Sn_T^2$ non-zero elements in $\bm Z$, which means that a fraction $1/n_S$ of the elements are non-zero. If $n_T=10, n_S=544$ (as in the simulation example to be considered later), then 0.1838 \% of the elements are non-zero. If we however consider using the temporal constraints $\bm A_1=\bm A_{T,1}$, then $1/n_T=10 \%$ of the elements of the projector matrix $\bm Z$ is non-zero. We observe a striking difference. This affects the run time.

For the SC-constraints, we have the additional constraints $[\bm d\otimes \bm I_{n_S}]\bm\delta=\bm 0$. These can be incorporated to the projection part as well by making these orthogonal to the constraints already included, as well as scaling them to have norm 1. Assuming $\bm A_1=\bm A_{1,S}$, this can be achieved by standard scaling and centering of $\bm d$, that is 
\[
\tilde d_t=\frac{d_t-\bar d}{\sqrt{n_T^{-1}\sum_{i=1}^{n_T}(d_i-\bar d)^2}}.
\]
We then can use $\bm A_1$ as the combination of $\bm A_{S,1}$ and $\bm [\tilde{\bm d}\otimes \bm I_{n_S}]$.

\subsection{Does the mixed effect and kriging methods give the same posterior?}

The mixed effect approach is used to the correct the prior while the kriging method is used to correct the distribution during the execution of the algorithm. We now show that they provide the same inference. Assuming $\bm A\bm x=\bm 0$, we have
\begin{align*}
p(\bm x|\bm A\bm x=\bm 0,\bm y)
=&\frac{p(\bm x)p(\bm Ax=\bm 0|\bm x)p(\bm y|\bm x,\bm A\bm x=\bm 0)}
{p(\bm y|\bm A\bm x=\bm 0)p(\bm A\bm x=\bm 0)}\\
=&\frac{p(\bm x)p(\bm y|\bm x)}
{p(\bm y|\bm A\bm x=\bm 0)p(\bm A\bm x=\bm 0)}\\
=&\frac{p(\bm x|\bm A\bm x=\bm 0)p(\bm y|\bm x)}
{p(\bm y)},
\end{align*} 
where we in the second equality have used that $p(\bm Ax=\bm 0|\bm x)=1$ when $\bm A\bm x=\bm 0$ and that given $\bm x$, the extra conditioning $\bm A\bm x=\bm 0$ does not provide extra information.
In the last equality we have used that
$p(\bm y|\bm A\bm x=\bm 0)$ reduce to $p(\bm y)$ assuming that the constraints imposed only are related to identifiability so that the density of $\bm y$ do not depend on these constraints. From this we see that conditioning in the posterior is equivalent to conditioning in the prior.


\section{Numerical studies}{\label{sec:numstud}}

In this section we perform numerical studies in order to compare the HyMiK method with the standard conditioning by kriging method. Three data sets are considered. The first is a simulated data set where we use a neighbor structure based on districts within Germany. The second example is a real data set of dengue fewer cases in Brazil, previously studied in \cite{lowe2021combined}. The  last example is a real data set consisting of cases of positive tests of COVID-19 in Norway. 

The computational gain is measured as the fraction between computing time using the standard conditioning by kriging method and the time for the HyMiK method. We call this the \emph{computational factor}.  

All computations were performed on a linux server (kernel 5.4.0-1100.azure)  using INLA version 22.05.07 with the Pardiso-extension~\citep{van2021NewFrontiers}. The Pardiso extension uses a modern sparse-matrix library, which gives faster inference for complex models. A maximum number of 10 threads were used in all calls to INLA.

All datasets and scripts are available at the github repository \url{https://github.com/geirstorvik/INLAconstraints}. 
When calculating marginal (log)likelihoods, care has to be taken due to that for some models the normalizing constant, i.e. the term $0.5\log(|\bm R|)$ (with $\bm R$ being the structural matrix taking constraints into account), is missing. Calculation of this term was performed as a post-processing step. Further, the interpretation of the marginal likelihoods for the GC constraints is somewhat difficult due to that the priors are still improper under the GC constraints. The prior is first made proper by considering $\bm Q_\varepsilon$, see \eqref{eq:Q_eps}. The constraints only make the posterior proper and not the prior, which have some directions with very large variances. Technically, the variance in some directions  is still $1/\varepsilon$ under the constraints, which gives what we call an improper prior. For the SC constraints, the priors are proper under the constraints (where the additional constraint $\sum t\alpha_t$ is needed on the temporal term). We still consider $\bm Q_\varepsilon$ but the constraints make the variance in the direction which originally had variance $1/\varepsilon$ now 0. This distribution is singular, but proper.    
\subsection{Simulated data}\label{sec:res.sim}

We simulated data from  model~\eqref{eq:model} with $p_y$ corresponding to the Poisson model with log-link. The spatial structure was based on the neighbor structure of the $n_S=544$ districts of Germany.  This neighbor structure is available in the INLA package through the Germany data, see e.g. \href{https://rdrr.io/github/inbo/INLA/src/demo/Graph.R}{https://rdrr.io/github/inbo/INLA/src/demo/Graph.R}. This neighbor structure was used to define the precision matrix for the main spatial random effect through the ICAR model~\eqref{eq:ICAR}. For the temporal effect, we used the RW(2) model and the interaction term followed the type IV model of~\citet{knorr2000bayesian}, that is equation~\eqref{eq:Q.delta}. When simulating data, the SC constraints were used to obtain a proper prior of the interaction effect and also appropriate constraints where used to obtain proper main spatial and temporal effects. A summary of the procedure is given in the supplementary material, section~\ref{sec:sim.proc}.

For inference, we use the true model but consider either the GC or the SC constraints. 
Table~\ref{tab:Germany_sim_GC} shows the estimates of the hyperparameters and the fixed effect for the two sets of constraints while Figure~\ref{fig:Sim_Mean} compare the posterior mean of the interaction random effects for GC and SC, respectively. Figure~\ref{fig:Sim_SD} in the appendix shows the posterior standard deviations.
Exact correspondence is not to be expected, which has been explained in Section ~\ref{sec:projection method for inference}. Even though the new approach introduce an extra approximative $\kappa$
term (adding the value to get a proper normal distribution),  the new method is expected to be more accurate.
We do however see that there is a high correspondence between the two methods: the posterior means of the random effects are very similar, while the uncertainties (Figure~\ref{fig:Sim_SD}) are somewhat lower for the HyMiK method. The most striking difference between the parameter estimates is the estimate of $\tau_\delta$ which is somewhat larger for the standard method compared to the HyMiK method. This is valid for GC constraints only. 
We discuss this and computing times in Section ~\ref{sec:discussion on results}.

\begin{table}[t]
\centering
\begin{tabular}{|c|rr|rr||rr|rr|}
  \hline
  &\multicolumn{4}{c||}{GC}&\multicolumn{4}{c|}{SC}\\
  &\multicolumn{2}{c|}{Standard}&\multicolumn{2}{c||}{HyMiK}
  &\multicolumn{2}{c|}{Standard}&\multicolumn{2}{c|}{HyMiK}\\
 & mean & sd & mean & sd & mean & sd & mean & sd \\ 
   \hline
$\mu$ & 1.50 & 0.00 & 1.50 & 0.00 & 1.50 & 0.00 & 1.50 & 0.00 \\
$\tau_\alpha$  & 61.29 & 27.36 & 58.45 & 26.21 & 57.08 & 25.10 & 57.46 & 25.44 \\ 
$\tau_\gamma$  & 9.71 & 0.62 & 9.70 & 0.61 & 9.70 & 0.61 & 9.70 & 0.61 \\ 
$\tau_\delta$  & 24.99 & 0.92 & 20.82 & 0.82 & 17.07 & 0.66 & 17.07 & 0.66 \\
   \hline
Av. marg.lik&\multicolumn{2}{c|}{-2.184}&\multicolumn{2}{c||}{-2.183}&\multicolumn{2}{c|}{-2.165}&\multicolumn{2}{c|}{-2.183}\\
Comp. time&\multicolumn{2}{c|}{37 sec}&\multicolumn{2}{c||}{26 sec}&\multicolumn{2}{c|}{93 sec}&\multicolumn{2}{c|}{24 sec}\\
\hline
\end{tabular}
\caption{\label{tab:Germany_sim_GC}Simulated data: Estimates of the intercept and the hyperparameters based on the GC and the SC constraints for both the standard (conditioning by kriging) and the new HyMiK approach.}
\end{table}

\begin{figure}
\centering
\begin{tabular}{cc}
\includegraphics[width=0.4\textwidth,trim=2.5cm 0 2.5cm 0.38cm, clip]{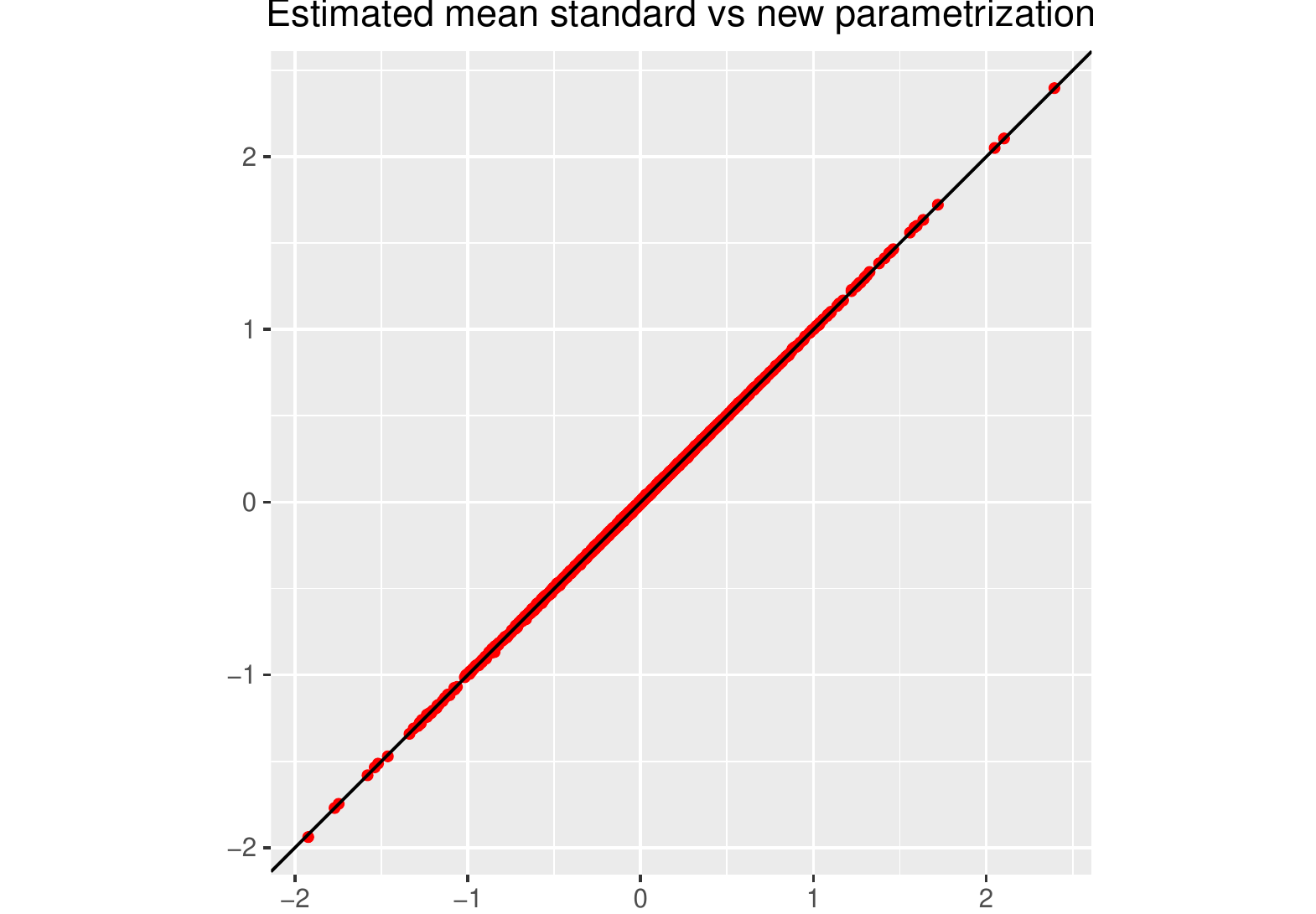}&
\includegraphics[width=0.4\textwidth,trim=2.5cm 0 2.5cm 0.38cm, clip]{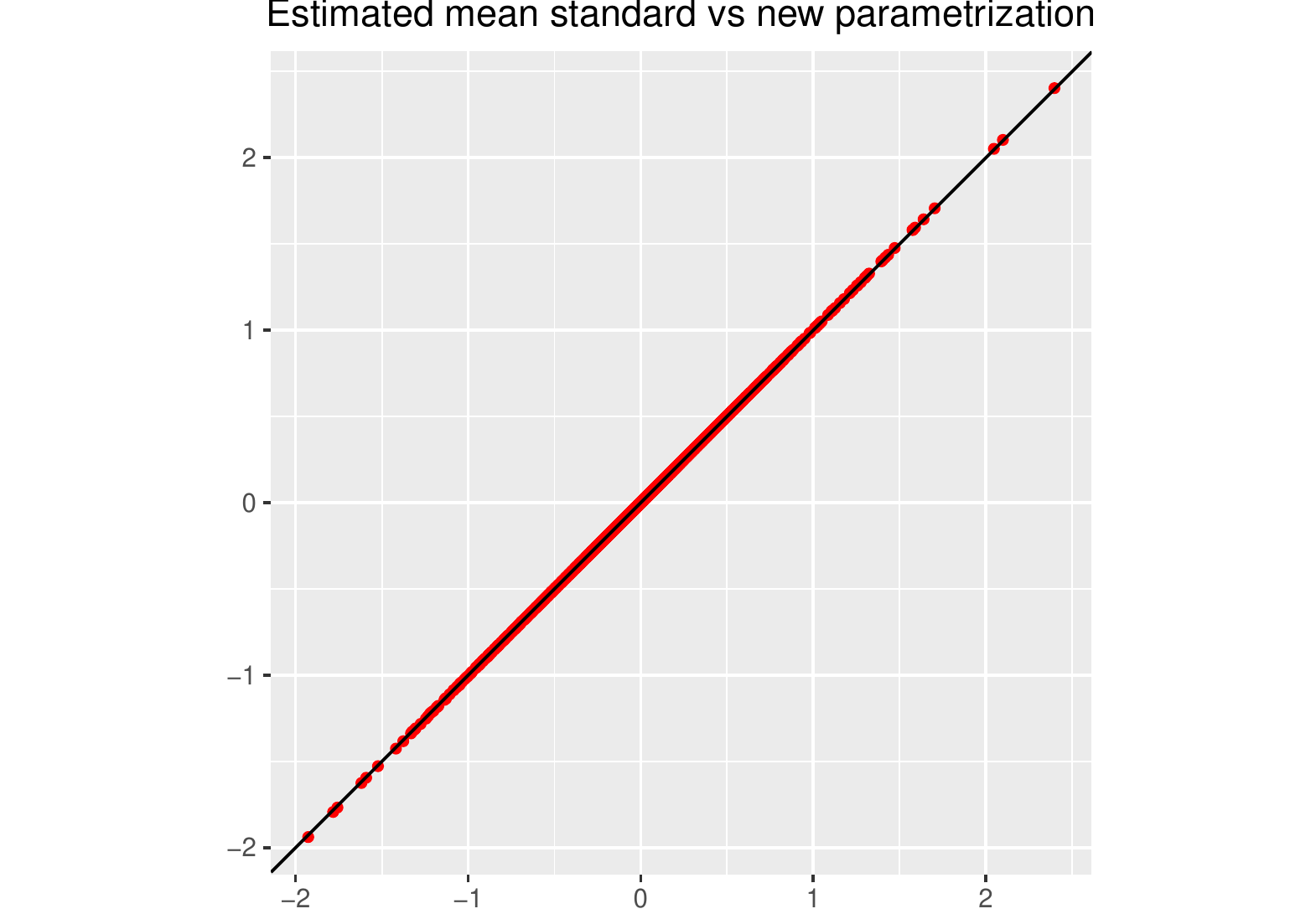}
\end{tabular}
\caption{Simulated data: Comparison of the posterior means between the standard method ($x$-axis) and the HyMiK approach ($y$-axis) for the interaction effects, based on the GC (left) and the SC (right) constraints. Standard deviations are shown in Figure~\ref{fig:Sim_SD} in the appendix.}
\label{fig:Sim_Mean}
\end{figure}

\subsection{Dengue case study}

In \cite{lowe2021combined} a spatio temporal model is fitted on data recording dengue fewer numbers between 2000 and 2019 given for the 558 microregions in Brazil. Several covariates and models are tested. The authors use distributed lag-non-linear models \citep{gasparrini2014modeling}. Essentially, the model is given by
\begin{equation}
y_{t,s}| \bm{\eta}\stackrel{ind}\sim\text{Neg-Binom}(E_s\exp(\eta_{t,s}))
\end{equation}
with $\eta_{t,s}$ following the structure of equation~\eqref{eq:model.eta} where the covariates $\bm u_{t,s}$ contain several derived features based on different risk factors at different time-lags (we have chosen a total of 10 features, somewhat smaller than used in the original work). The reader is referred to \citet{gasparrini2014modeling,gasparrini2017penalized} and the scripts within our GitHub repository for details.
Further, $E_s$ is the population in microregion $s$ divided by $100\,000$, so that response is equivalent to an incidence rate per 100\,000 people. 

From Figure~\ref{fig:Dengue_mean} we observe very good correspondence for both the GC and SC constraints. We observe, however, from Figure~\ref{fig:Dengue_sd}
that the standard errors of HyMiK parametrization are somewhat smaller than for the original parametrization. The estimates of the regression parameters are given in Table~\ref{tab:dengue.regr}, where we observe very good correspondence. The estimates of the hyperparameters and the intercept are given in Table~\ref{tab:dengue.hyper}. We observe very good correspondence, except for $\tau_\delta$ for the GC constraints, where there is some deviation. There is also a relatively large deviation for the $\tau_\delta$ parameter when estimating with SC constraints. We provide a general discussion of this and also compare computing times in Section \ref{sec:discussion on results}.

\begin{figure}
\centering
\begin{tabular}{cc}
\includegraphics[width=0.4\textwidth,trim=0.52cm 0.55cm 0 0.55cm, clip]{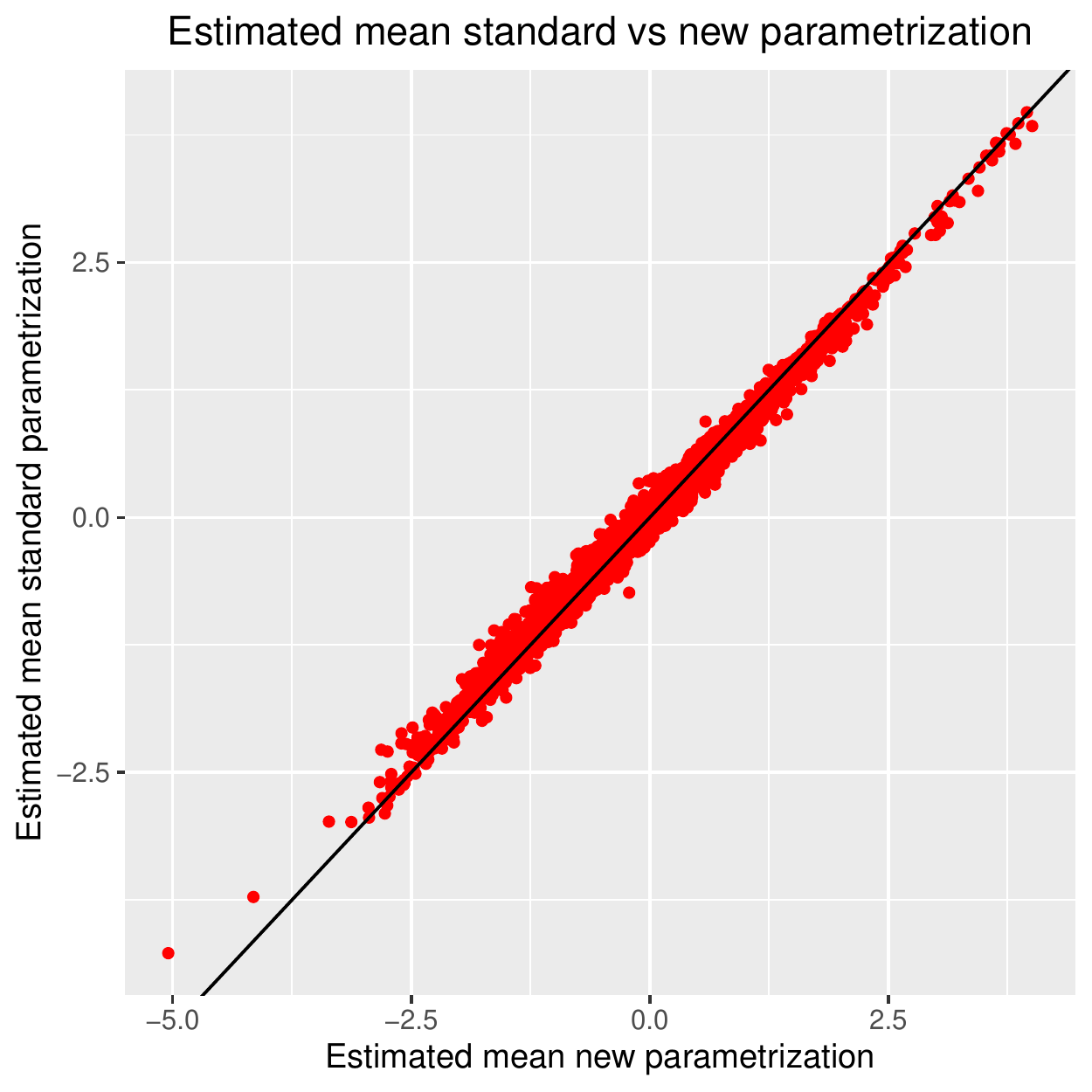}&
\includegraphics[width=0.4\textwidth,trim=0.52cm 0.55cm 0 0.55cm, clip]{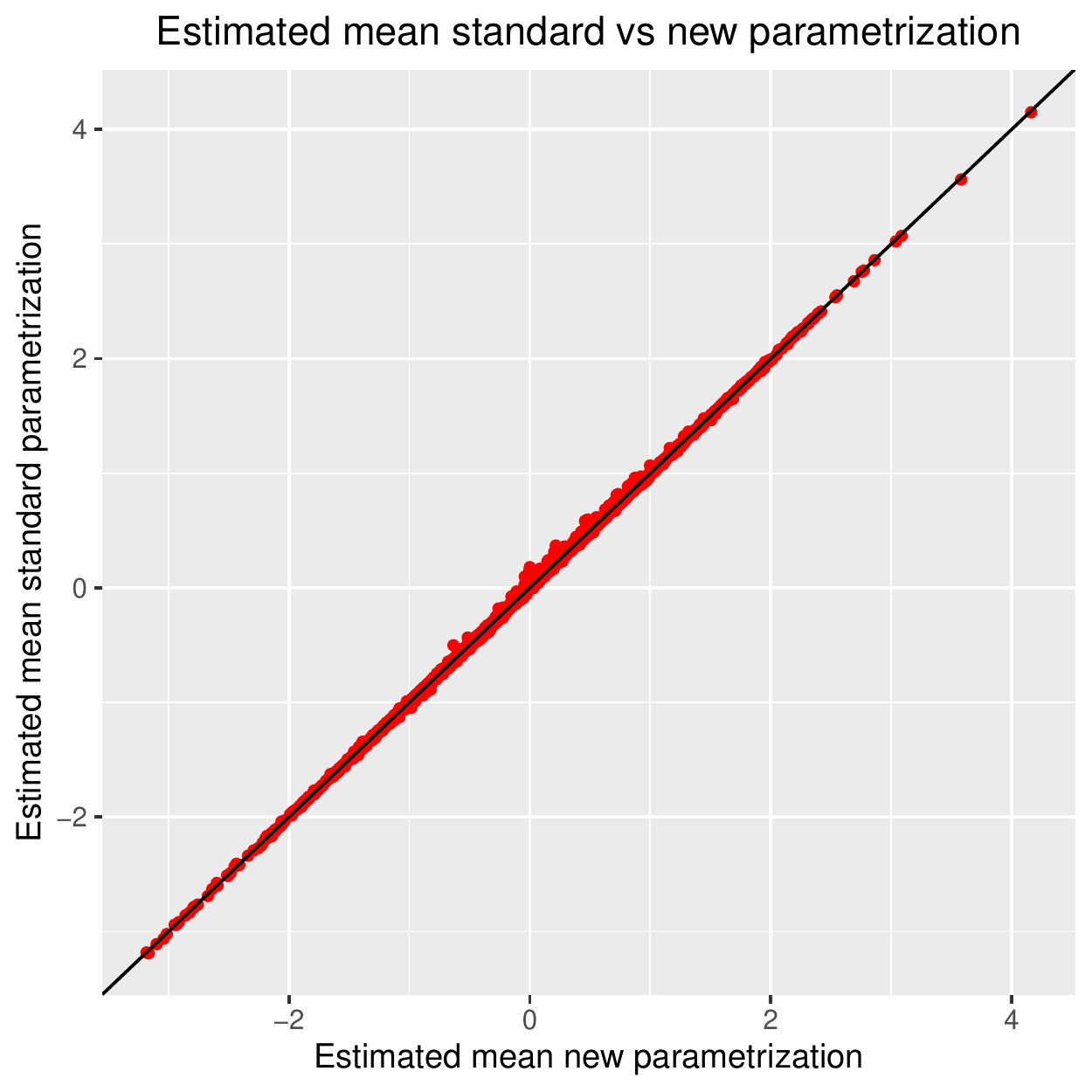}
\end{tabular}
\caption{Dengue data: Comparison of the posterior means between the standard method and the new approach for the interaction effects, based on the GC (left) and the SC (right) constraints. Standard deviations are shown in Figure~\ref{fig:Dengue_sd} in the appendix.}
\label{fig:Dengue_mean}
\end{figure}

\begin{table}[ht]
\centering
\begin{tabular}{|r|rr|rr||rr|rr|}
  \hline
  &\multicolumn{4}{c||}{GC}&\multicolumn{4}{c|}{SC}\\
  &\multicolumn{2}{c|}{Standard}&\multicolumn{2}{c||}{HyMiK}
  &\multicolumn{2}{c|}{Standard}&\multicolumn{2}{c|}{HyMiK}\\
  \hline
 & mean & sd & mean & sd & mean & sd & mean & sd \\ 
  \hline
$\mu$ &  -1.732 & 0.708 & -1.844 & 0.708 & -0.130 & 9.589 & -0.013 & 9.924 \\  
1/Disper &  0.860 & 0.008 & 0.882 & 0.009 & 0.724 & 0.009 & 0.725 & 0.006 \\ 
 $\tau_\alpha$ &5.957 & 1.129 & 5.916 & 0.997 & 11.592 & 1.550 & 2.637 & 0.630 \\
 $\tau_\gamma$  & 0.241 & 0.016 & 0.236 & 0.019 & 0.302 & 0.068 & 0.254 & 0.017 \\ 
 $\tau_\delta$ & 0.231 & 0.013 & 0.114 & 0.006 & 0.172 & 0.007 & 0.175 & 0.007 \\
     \hline
Av. marg.lik&\multicolumn{2}{c|}{-2.864}&\multicolumn{2}{c||}{-2.835}&\multicolumn{2}{c|}{-2.837}&\multicolumn{2}{c|}{-2.837}\\
   Comput.time&\multicolumn{2}{c|}{187 sec}&\multicolumn{2}{c||}{103 sec}&\multicolumn{2}{c|}{386 sec}&\multicolumn{2}{c|}{101 sec}\\
\hline
\end{tabular}
\caption{\label{tab:dengue.hyper}Estimates of intercept and hyperparameters for the Dengue data based on the GC and the SC constraints for both the standard and the new HyMiK approach. Estimates of the remaining regression coefficients are given in Table~\ref{tab:dengue.regr}.}
\end{table}

\subsection{Covid-19 data}
Here we consider the daily numbers of positive tests within each county in Norway during the Covid-19 pandemic. The data are available at \url{https://github.com/folkehelseinstituttet/surveillance_data/tree/master/covid19} and we have here considered data in the period 2020-09-02 to 2022-01-14 giving $n_T=500$ days with counts available for each of the $n_S=11$ counties in Norway. These data are given as cumulative numbers over time, but with some inconsistencies in that in a few cases the cumulative numbers are lower than the previous day. In those cases we have put the daily numbers to zero.
We assume the model
\begin{align*}
y_{t,s}|\bm\eta\stackrel{ind}{\sim}\text{Poisson}(E_{s}\exp(\eta_{t,s}))
\end{align*}
with $\eta_{t,s}$ given in~\eqref{eq:model} without any covariates. Here $E_s$ is the population size within county $s$, assumed to be constant within the time period considered. Figure \ref{fig:Covid_mean} indicates that the means of the latent variables are very similar for both the GC and SC constraint. Also Figure \ref{fig:Covid_sd} indicates very good correspondence between the standard and the HyMiK approaches for making inference in the Knorr Held interaction models for both GC and SC constraints. The estimates for the hyperparameters, Table~\ref{tab:covid.hyper}, are also very similar, except for the precision parameters for the interaction effects, valid when estimating with GC constraints and also SC constraints. See Section \ref{sec:discussion on results} for a discussion.

\begin{figure}
\centering
\begin{tabular}{cc}
\includegraphics[width=0.4\textwidth,trim=0.52cm 0.55cm 0 0.55cm, clip]{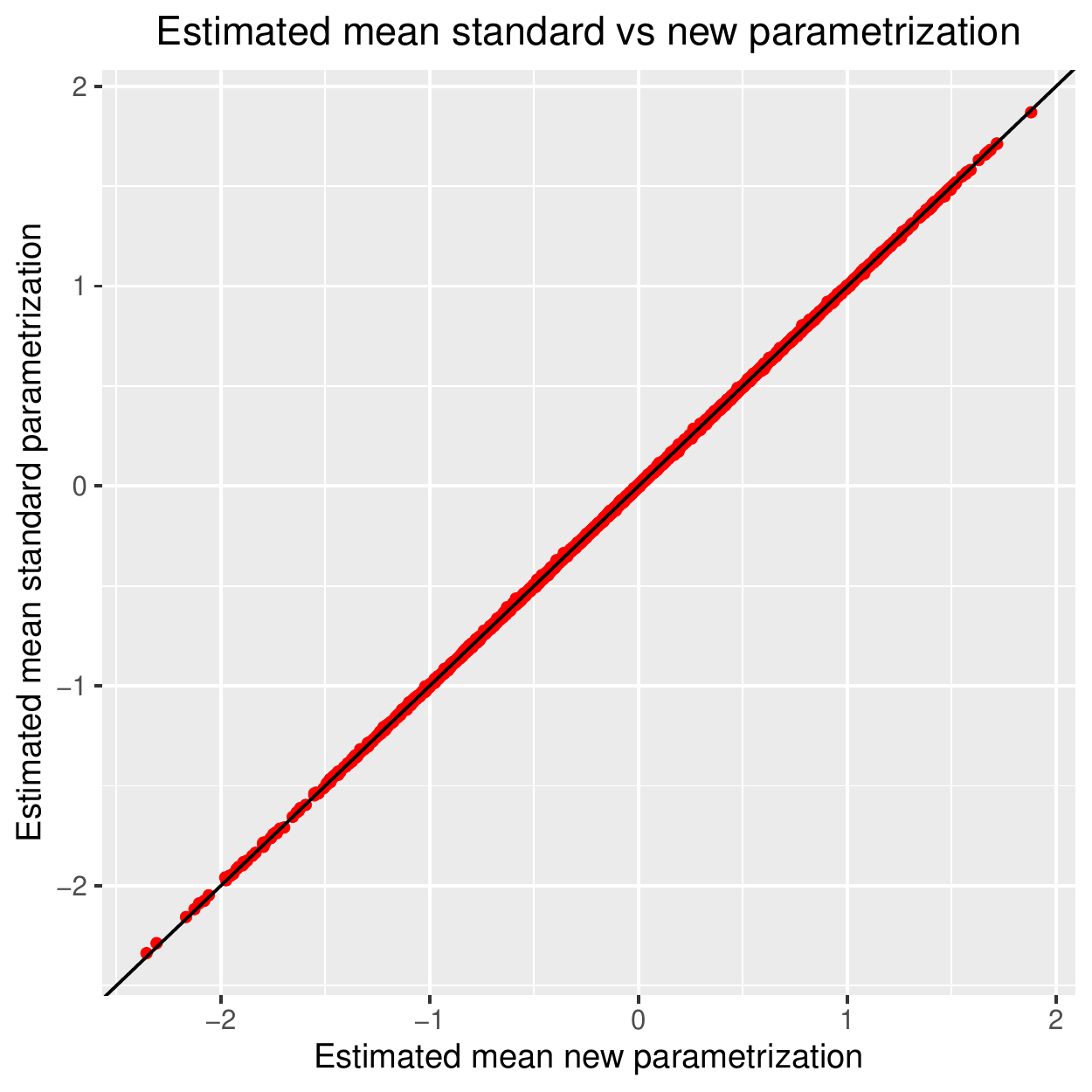}&
\includegraphics[width=0.4\textwidth,trim=0.52cm 0.55cm 0 0.55cm, clip]{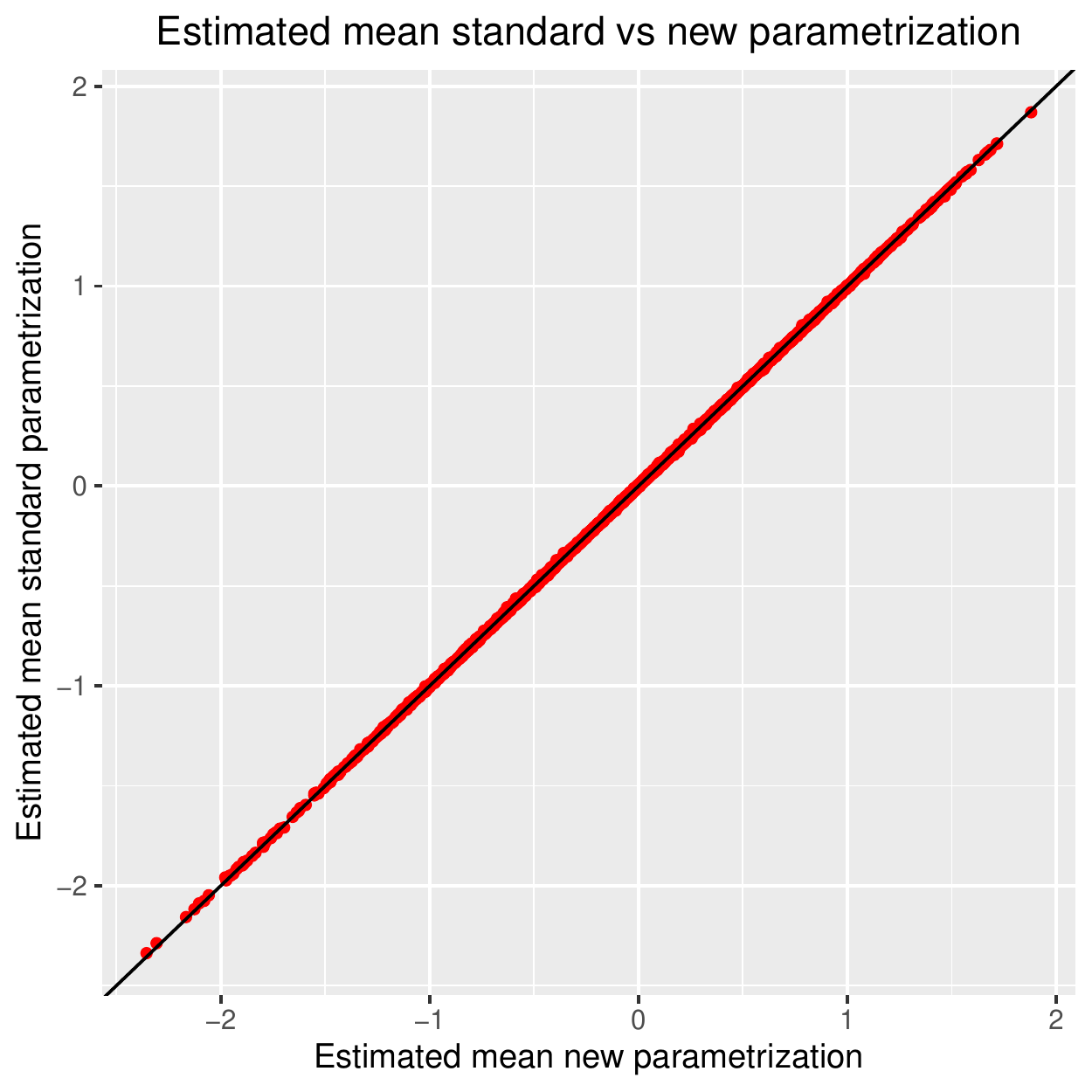}
\end{tabular}
\caption{Covid data: Comparison of the posterior means between the standard method (x-axis) and the new HyMiK approach (y-axis) for the interaction effects, based on the GC (left) and the SC (right) constraints. Standard deviations are shown in Figure~\ref{fig:Covid_sd} in the appendix.}
\label{fig:Covid_mean}
\end{figure}

\begin{table}[ht]
\centering
\begin{tabular}{|r|rr|rr||rr|rr|}
  \hline
  &\multicolumn{4}{c||}{GC}&\multicolumn{4}{c|}{SC}\\
  &\multicolumn{2}{c|}{Standard}&\multicolumn{2}{c||}{HyMiK}
  &\multicolumn{2}{c|}{Standard}&\multicolumn{2}{c|}{HyMiK}\\
  \hline
 & mean & sd & mean & sd & mean & sd & mean & sd \\ 
  \hline
$\mu$ &-9.421 & 0.004 & -9.421 & 0.004 & -9.421 & 0.004 & -9.421 & 0.004 \\ 
$\tau_\alpha$ & 9.522 & 0.706 & 9.521 & 0.711 & 9.528 & 0.711 & 9.498 & 0.696 \\ 
$\tau_\gamma$  &  5.210 & 2.039 & 5.408 & 2.185 & 5.334 & 1.989 & 5.299 & 2.065 \\ 
$\tau_\delta$  & 12.487 & 0.577 & 11.512 & 0.558 & 12.437 & 0.594 & 11.568 & 0.538 \\ 
   \hline
Av. marg.lik&\multicolumn{2}{c|}{-4.338}&\multicolumn{2}{c||}{-4.329}&\multicolumn{2}{c|}{-4.338}&\multicolumn{2}{c|}{-4.329}\\
   Comput.time&\multicolumn{2}{c|}{31 sec}&\multicolumn{2}{c||}{10 sec}&\multicolumn{2}{c|}{30 sec}&\multicolumn{2}{c|}{9 sec}\\
\hline
\end{tabular}
\caption{\label{tab:covid.hyper}Estimates of intercept and hyperparameters for the Covid data based on the GC and the SC constraints for both the standardS and the new HyMiK approach.}
\end{table}

\subsection{Discussion on results}
\label{sec:discussion on results}

We first consider the simulated and dengue fever datasets. Here $n_S\gg n_T$. With the HyMiK method we can handle 
$n_S$ of the $n_S+n_T-1$ GC constraints through the mixed effect model. For the
SC constraints, $2n_S$ out of $2n_S+n_T-1$ constraints can be considered within the mixed effect model.
The remaining constraints $[\bm I_{n_T}\otimes\bm 1_{s_N}^T]\bm\delta=\bm 0_{n_T}$ need to be enforced through conditioning by kriging. 
This means the number of time steps should not be too large. When considering the Covid-19 dataset, where $n_T\gg n_S$, we can remove $n_T$ constraints from either set of constraints. Conditioning by kriging is then used on the last constraints, a relative low number due to that $n_S=11$ in this case.

The main difference between the dengue and simulated data cases and the Covid-19 data, is the number of spatial and temporal regions. In the former cases, we have many more regions than temporal time points and in the latter case the opposite is true. This means we should take different approaches when estimating the models. When $n_S\gg n_T$ we should take the approach used in the dengue and simulated data cases, and when $n_T\gg n_S$ we should use same approach as in the Covid-19 data problem, 

The main benefit of the HyMiK method is the reduced run time of the model. We observe that the computational factor is 3 for both the GC constraints and the SC constraints when we consider the COVID data problem. When considering the dengue dataset the computational factors are 1.8 and 3.8, where the factor is greatest when using the SC constraint set. For the simulated dataset the factors are 1.4 and 3.9. We therefore in all examples obtain a significant reduction in computing time.  We also observe that the computing time is reduced the most for the SC constraints when we compare HyMiK and the original method, which is as expected since the SC constraints have the largest number of constraints in the original formulation, so that using the traditional kriging method in INLA should take the largest amount of time. The reduction in computing time for the large latent field in the dengue dataset problem is considerable and very fast inference can be obtained with the HyMik approach.
This is natural as several constraints are dropped in HyMiK.

Somewhat surprisingly, HyMiK ran a little slower with GC constraints on the COVID dataset and also the simulated dataset. This is not as expected. The reason is that the projection matrix is more  dense for the SC constrains, but the other constraints remain unchanged.

For all case studies, the precision parameters are the parameters with estimates that differ the most, somewhat surpricing given that the interaction terms themselves are very similar. We find it plausible that the HyMiK is more precise (as explained in Section \ref{sec:projection method for inference}), so that the estimates of the precision parameters for the new method should be more accurate.  

 One way to tell how much of the variance that is explained by each of the main effects and interaction effects is by considering the precision parameters. We have previously noted that the hyperparameters differ when comparing the HyMiK and the standard conditional by kriging method. We then get different results when quantifying the proportion of variance explained by the different effects. As previously explained, we expect the HyMiK method to be more exact, as the kriging step is done in a lower dimensional space, meaning we should consider the proportions by using the new method. 

In this paper we have chosen not to scale the precision matrices when performing inference~\citep[for scaling see][]{sorbye2014scaling}, as the focus is on demonstrating differences and similarities of HyMiK and the condition by kriging methods. However, some experiments where scaling was included gave similar correspondence between the two methods for dealing with constraints.

We have also computed the average marginal likelihood for the different models. As INLA does not include the determinant of $\bm Q$ for the \texttt{generic0} model, we had to calculate this as a post-processing step to obtain the correct average marginal likelihood (a routine for this correction is provided in the GitHub repository). From Tables~\ref{tab:Germany_sim_GC}, ~\ref{tab:dengue.hyper}  and~\ref{tab:covid.hyper} we see that the values are very similar. 

We have used the new HyMiK approach for estimation of models within INLA, which makes the model less dependent on the number of observations. The inference is still dependent on the size of the latent field. This is demonstrated in the simulation experiment, where the number of observations is 30 times the size of the linear predictor in the usual space time set up.

\section{Summary, general discussion and conclusion}
\label{section:Discussion}

 The idea behind the HyMiK method is simple: Instead of handling all the constraints by the conditioning by kriging approach, some of the constraints are built in to the model through a mixed model approach. The remaining constraints are handled by the conditioning by kriging approach, but now on a (much) lower number of constraints. INLA is used to estimate the posterior as it has the conditioning by kriging step and also the mixed model implemented. 

 Both the conditioning by kriging approach and the mixed model approach add small numbers to the diagonals of the precision matrices in order for the intrinsic models to become proper, which is needed for the numerical calculations to not break down. In addition, INLA is based on a Gaussian approximation for handling non-Gaussian data.
 Both methods thereby introduce some numerical errors. We have however argued that the use of the mixed model through the HyMiK approach should be numerically more accurate due to that corrections for some of the constraints are performed prior to the Gaussian approximation. 
 
 It is crucial to use the correct set of constraints when estimating the different types of Knorr Held models. \cite{fattah2022approximate} follow \cite{schrodle2011spatio} while Gocioa and coauthors have proposed the constraints as stated in \cite{goicoa2018spatio}. Both say that constraining the field with wrong constraint sets will give wrong estimates of model parameters. We can fit both set of constraints. The user has to decide which constraints to use. We note that while GC constraints may have less constraints, SC constraints will always lead to a proper space time prior and can thus be used with missing values present. Given the set of constraints, there is a design choice in which constraints that should be handled by the mixed effect approach and  which by conditioning by kriging. We have suggested two possibilities, which one to choose depends on the size of the temporal component compared to the spatial component.
 
 The recent paper \cite{OROZCOACOSTA2023107403} also considers estimating large spatio-temporal fields. The authors suggest to use a divide and conquer approach. The approach is approximate since the spatial domain is divided into smaller domains, where spatio-temporal models are fitted on the smaller spatial domains. This approach can benefit from our model reparametrization as well, as each submodel can utilize our HyMiK parametrization to obtain faster estimation. The user can choose to use our approach with INLA, which is more exact, or combine our reparametrization with \citet{OROZCOACOSTA2023107403}. Combining \citet{OROZCOACOSTA2023107403} with our reparametrization is further work. We have focused on making a new computational efficient procedure. As we can see, it can easily be combined with several other estimation procedures as well. As \citet{OROZCOACOSTA2023107403} also uses INLA, it should be relatively straightforward to combine the two methods. 

 When comparing our method to the method of \cite{fattah2022approximate}, called INLAPLUS, our method can also be used with GC constraints. This makes our method more flexible. Our method can also be used on severs and ordinary computers without vast number of cores available. We have found similar speedups, which means our method is the preferred one when there are limited computational resources available. However, INLAPLUS can give further reduction in computing time for SC constraints when computational resources are not an issue. 

HyMiK is faster than the traditional way of estimating constrained models in INLA, both for GC constraints and for SC constraints. We observe very close correspondence between estimated mean and standard deviations of the field, but that the hyperparameters can be somewhat different. Our method is expected to be more accurate, as previously explained. We have implemented procedures that can be used to set up the models easily. The preparation is fast. In total, we argue that HyMiK should become the standard procedure of estimating the Knorr Held type IV interaction model in INLA.  

\section*{Acknowledgment}

This research is partly funded by the Norwegian Research Council research-based innovation center \href{BigInsight.no}{BigInsight}, project no 237718.
  We thank Professor H{\aa}vard Rue for clarifying some issues related to how INLA is implemented. 
 All code is available at the GitHub repository \url{https://github.com/geirstorvik/INLAconstraints}.
 
\bibliographystyle{chicago}
\bibliography{references}

\newpage


\bigskip
\begin{center}
{\large\bf SUPPLEMENTARY MATERIAL}
\end{center}

\setcounter{section}{0}
\renewcommand{\thesection}{A-\arabic{section}}

\setcounter{figure}{0}
\renewcommand{\thefigure}{A-\arabic{figure}}
\setcounter{table}{0}
\renewcommand{\thetable}{A-\arabic{table}}
\setcounter{equation}{0}
\renewcommand{\theequation}{A\arabic{equation}}

\section{Description of procedure for constructing simulated data}\label{sec:sim.proc}

The simulated data in section~\ref{sec:res.sim} are generated by the following procedure:
\begin{itemize}
    \item Use the Germany graph to make $\bm Q_\alpha$.
    \item Let the temporal component be a random walk of order 2 with 10 time points
    \item Scale the precision matrices $\bm Q_\alpha$ and $\bm Q_\gamma$ to obtain scaled versions $\bm Q^*_\alpha$ and $\bm Q_\gamma^*$ and set $\bm Q_{\delta}^*=\bm Q^*_\alpha \otimes \bm Q^*_\gamma$. The function  \texttt{inla.scale.model} is used to scale the matrices.
    \item Generate one sample from each of the IGMRF models using the SC constraints on the interaction term and appropriate sum to zero constraints on the two remaining main effects. This means proper priors are used. The function \texttt{inla.qsample} is used to generate the samples.  
    \item Simulate 30 response variables from Poisson($\bm \eta_{t,s}$)for each linear predictor $\eta_{t,s}$. 
    \item Fit the models using INLA. 
\end{itemize}
Scaling of a precision matrix means it is divided by a constant. This constant is found by calculating a mean of the diagonal elements of the inverse of the precision matrix under suitable constraints. Then the ``typical'' marginal variance of the prior is 1. See e.g \cite{sorbye2014scaling} or \cite{riebler2016intuitive} for details.  
The script for simulating the data is available at our GitHub repository under the name \texttt{SimulateModel.R}.

\section{Additional results}
Here, figures showing the similarities between the posterior standard errors of the interaction terms for the simulated data, the dengue data and the Covid-19 data are shown in figures~\ref{fig:Sim_SD}, \ref{fig:Dengue_sd} and~\ref{fig:Covid_sd}, respectively. Further, Table~\ref{tab:dengue.regr} contain estimates of the regression coefficients for the dengue data set.
\begin{figure}
\centering
\begin{tabular}{cc}
\includegraphics[width=0.35\textwidth,trim=2.92cm 0.55cm 2.5cm 0.32cm, clip]{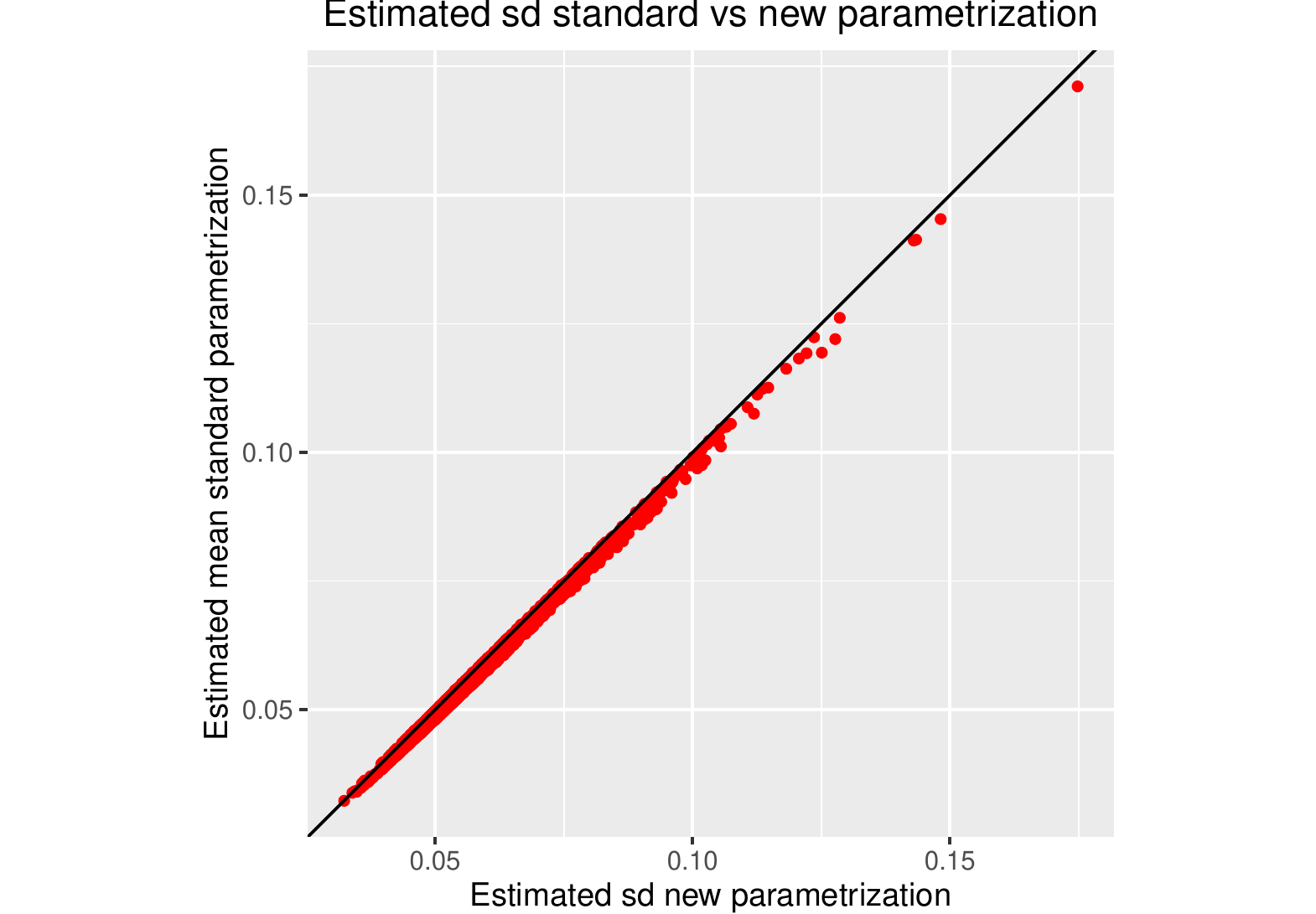}&
\includegraphics[width=0.35\textwidth,trim=2.92cm 0.55cm 2.5cm 0.32cm, clip]{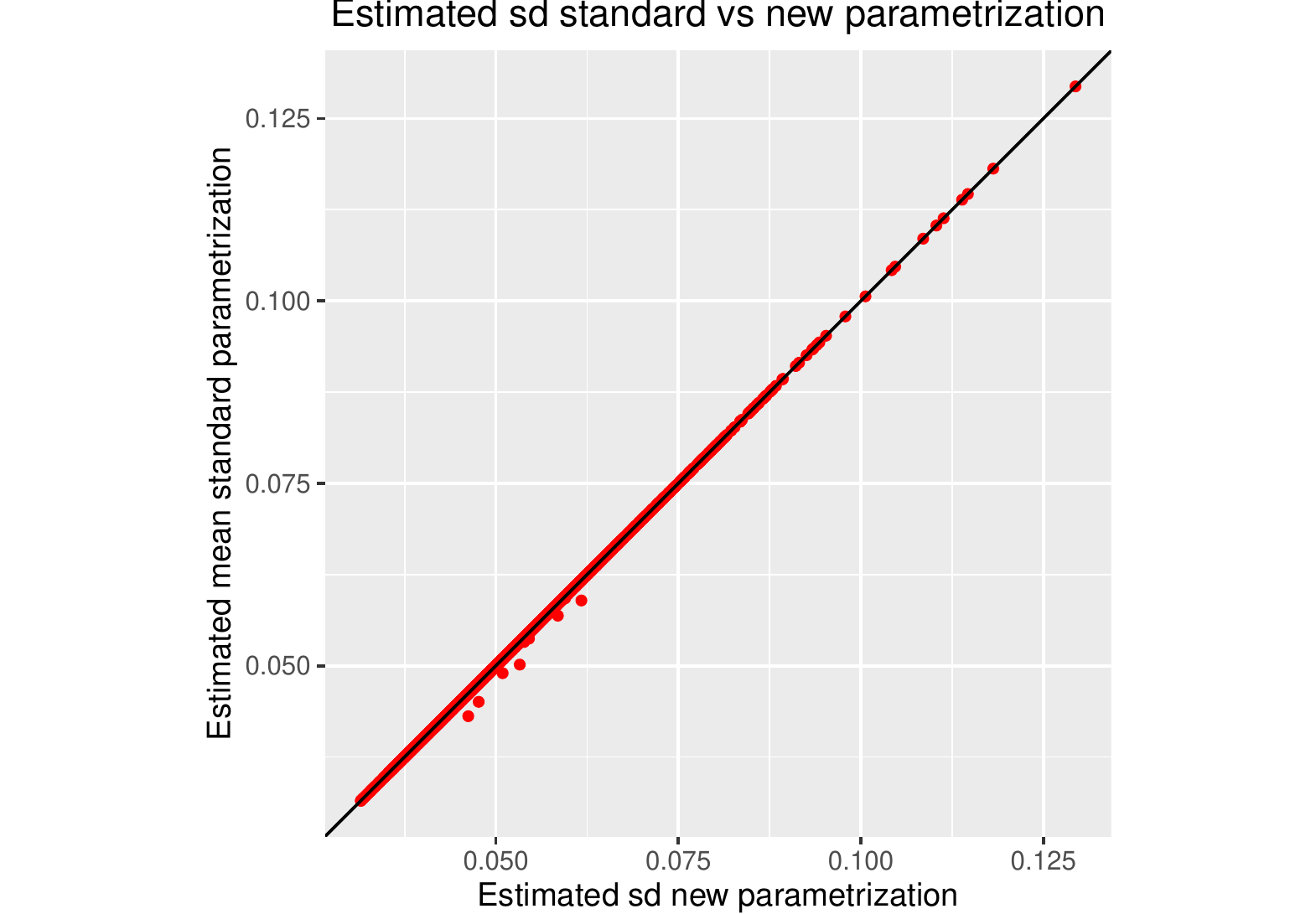}
\end{tabular}
\caption{Simulated data: Comparison of the posterior standard deviation between the standard method and the new approach  for the interaction effects, based on the GC (left) and the SC (right) constraints.}
\label{fig:Sim_SD}
\end{figure}

\begin{figure}
\centering
\begin{tabular}{cc}
\includegraphics[width=0.35\textwidth,trim=0.52cm 0.55cm 0 0.55cm, clip]{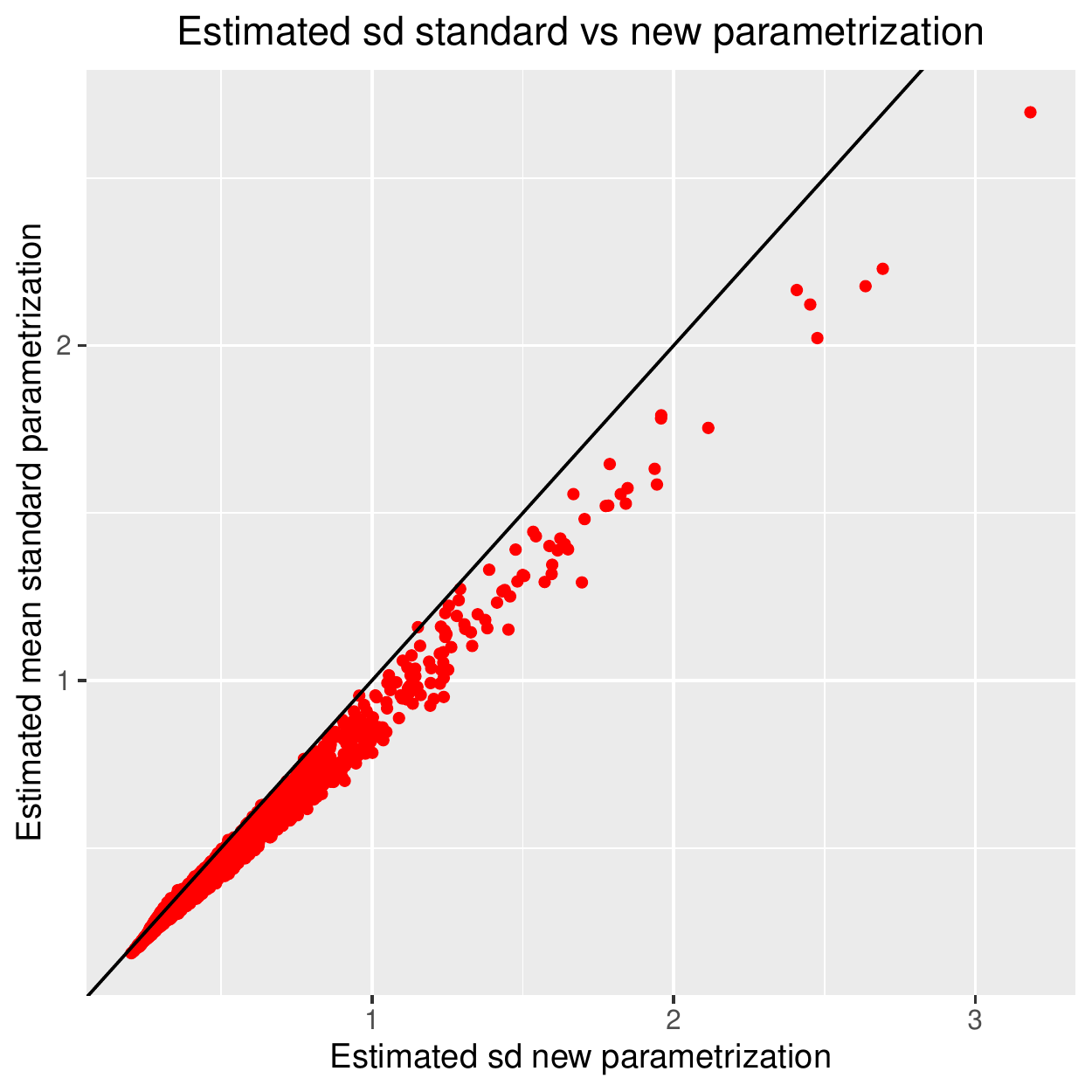}&
\includegraphics[width=0.35\textwidth,trim=0.52cm 0.55cm 0 0.55cm, clip]{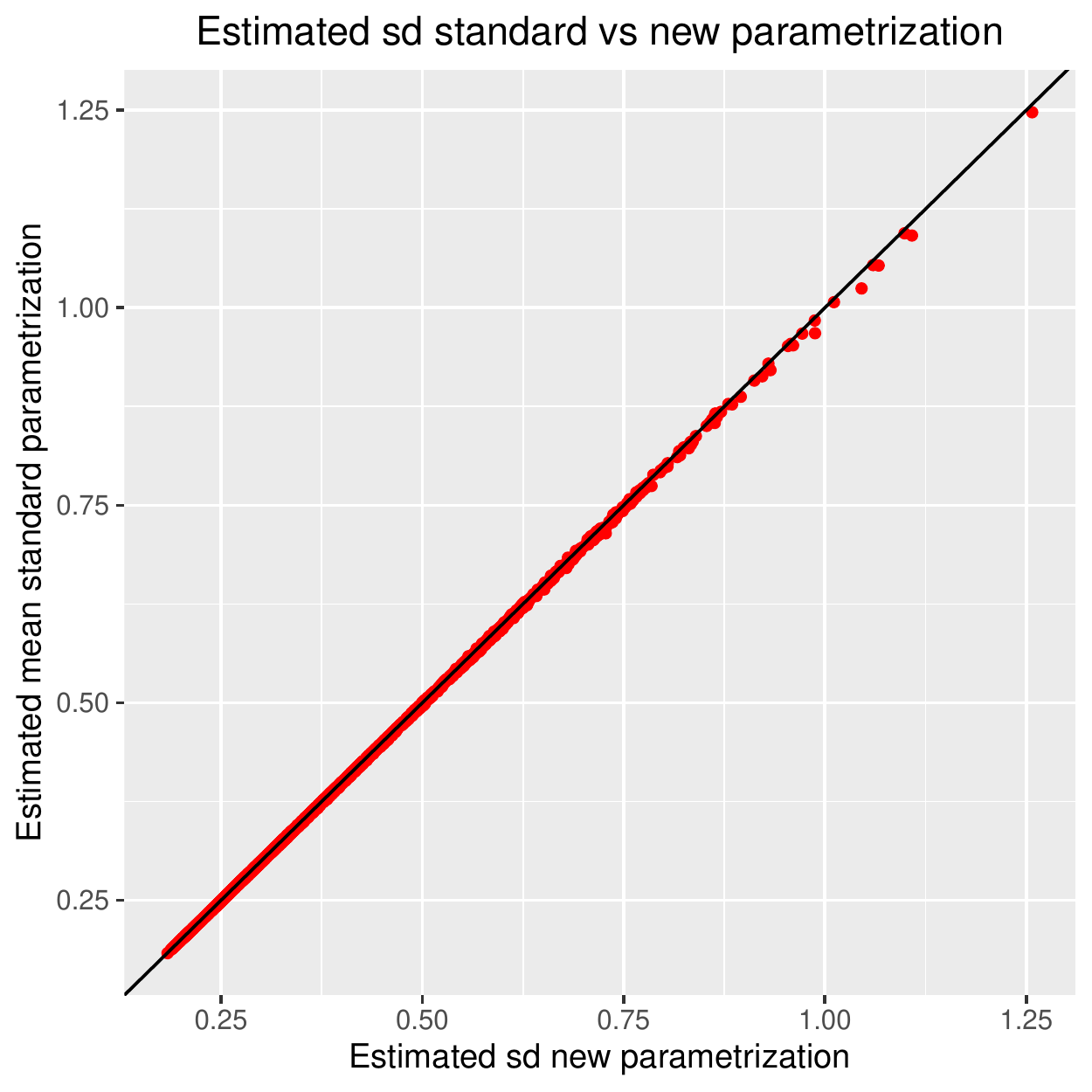}
\end{tabular}
\caption{Dengue data: Comparison of the posterior standard deviation between the standard method and the new approach  for the interaction effects, based on the GC (left) and the SC (right) constraints.}
\label{fig:Dengue_sd}
\end{figure}

\begin{table}[ht]
\centering
\begin{tabular}{|r|rrrr||rrrr|}
   \hline
  &\multicolumn{4}{c||}{GC}&\multicolumn{4}{c|}{SC}\\
  &\multicolumn{2}{c|}{Standard}&\multicolumn{2}{c||}{HyMiK}
  &\multicolumn{2}{c|}{Standard}&\multicolumn{2}{c|}{HyMiK}\\
  \hline
 & mean & sd & mean & sd & mean & sd & mean & sd \\ 
  \hline
  $\beta_1$ & 0.029 & 0.003 & 0.029 & 0.003 & 0.027 & 0.003 & 0.027 & 0.003 \\ 
  $\beta_2$ & -0.121 & 0.182 & -0.091 & 0.181 & -0.254 & 0.185 & -0.239 & 0.185 \\ 
  $\beta_3$ & 1.048 & 0.147 & 1.033 & 0.147 & 1.031 & 0.150 & 1.019 & 0.151 \\ 
  $\beta_4$ & -0.568 & 0.115 & -0.557 & 0.115 & -0.412 & 0.118 & -0.424 & 0.119 \\ 
  $\beta_5$ & 1.201 & 0.571 & 1.277 & 0.570 & 0.776 & 0.583 & 0.793 & 0.584 \\ 
  $\beta_6$ & -1.476 & 0.466 & -1.352 & 0.466 & -1.293 & 0.477 & -1.307 & 0.478 \\ 
  $\beta_7$ & -0.321 & 0.372 & -0.294 & 0.372 & -0.036 & 0.385 & -0.062 & 0.386 \\ 
  $\beta_8$ & 1.953 & 0.135 & 1.993 & 0.135 & 1.974 & 0.139 & 2.001 & 0.139 \\ 
  $\beta_9$ & 2.465 & 0.109 & 2.419 & 0.108 & 2.346 & 0.111 & 2.321 & 0.111 \\ 
  $\beta_{10}$ & -1.769 & 0.081 & -1.767 & 0.081 & -1.843 & 0.085 & -1.858 & 0.085 \\ 
\hline
 \end{tabular}
\caption{\label{tab:dengue.regr}Estimates of regression coefficients for the Dengue data.}
\end{table}

\begin{figure}
\centering
\begin{tabular}{cc}
\includegraphics[width=0.35\textwidth,trim=0.92cm 0.6cm 0 0.6cm, clip]{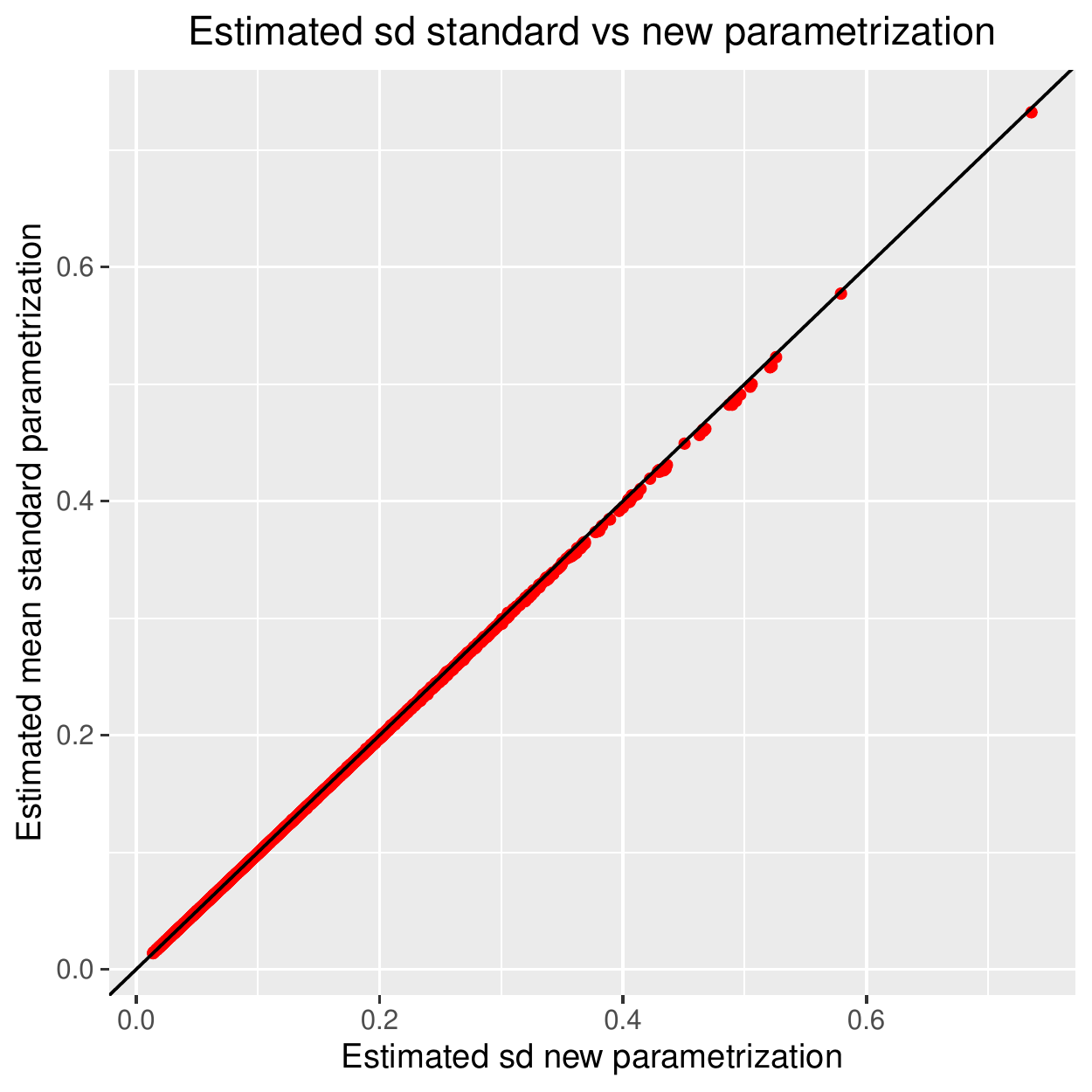}&
\includegraphics[width=0.35\textwidth,trim=0.92cm 0.6cm 0 0.6cm, clip]{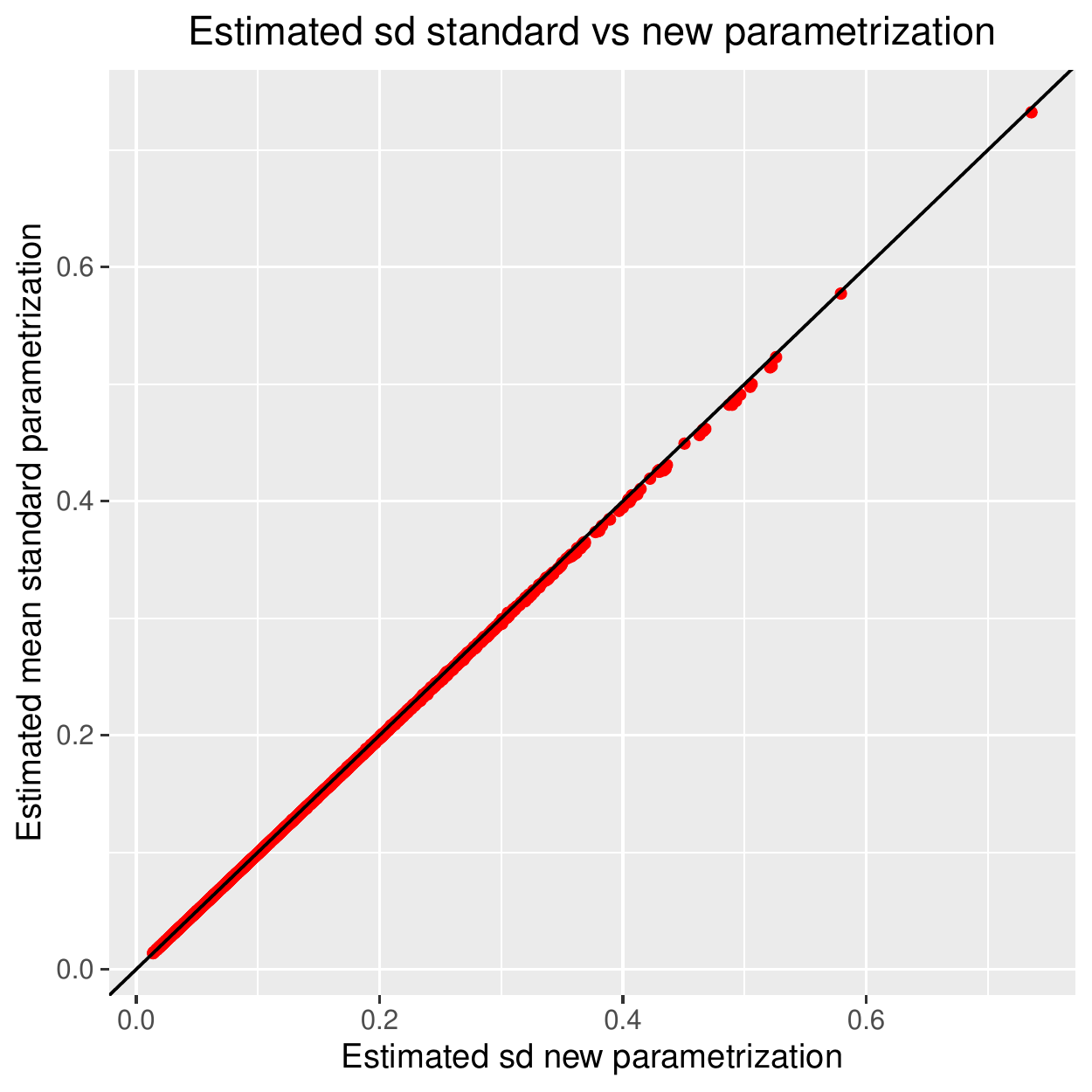}
\end{tabular}
\caption{Covid data: Comparison of the posterior standard deviation between the standard method and the new approach  for the interaction effects, based on the GC (left) and the SC (right) constraints.}
\label{fig:Covid_sd}
\end{figure}

-